\documentstyle[prl,eqsecnum,aps]{revtex}
\twocolumn

\begin{document}
\author{Jian Qi Shen\footnote{E-mail address: jqshen@coer.zju.edu.cn}}
\address{Zhejiang Institute of Modern Physics and Department of Physics,
Zhejiang University, Hangzhou 310027, P.R. China}
\date{\today}
\title{Comoving suppression mechanism and cosmological constant problem}
\maketitle

\begin{abstract}
In this paper, we assume that the observer is fixed in a comoving
frame of reference with $g_{00}=\frac{\lambda}{\Lambda}$, where
$\lambda$ and $\Lambda$ denote the comoving parameter and the
cosmological constant, respectively. By using the {\it comoving
suppression mechanism} and {\it Mach's principle} (the latter of
which is used to determine the comoving parameter $\lambda$), we
calculate the vacuum energy density of quantum fluctuation field
in the above-mentioned comoving frame of reference. It is shown
that in such a comoving frame of reference, the cosmological
constant will greatly decrease by many orders of magnitude (if
Mach's principle is applied to this calculation, then it will be
shown that $\Lambda$ is reduced by about 120 orders of magnitude).
Additionally, we briefly discuss the related topics such as the
varying observed speed of light ($\frac{{\rm d}c}{{\rm
d}t}={\mathcal O}\left(10^{-9}{\rm m/s}^{2}\right)$) and the
mystery of anomalous acceleration ($\sim 10^{-9}{\rm m/s}^{2}$)
acquired by the Pioneer 10/11, Galileo and Ulysses spacecrafts.

\end{abstract}

\section{Introduction}
Historically, the problem of vacuum energy density or cosmological
constant\cite{history} has been considered for many years. There
are now two cosmological constant problems. One is that why the
observed value of the vacuum energy density is so small (the ratio
of experimental value to the theoretical one is only $10^{-120}$).
The other one is to understand why the observed vacuum energy
density is not only small, but also, as current Type Ia supernova
observations seem to indicate, of the same order of magnitude as
the present mass density of the universe\cite{Weinberg}. It is
readily verified that the quantum vacuum fluctuation energy
density of one field is $U=\frac{c^{7}}{8\pi^{2}\hbar G^{2}}$ and
the cosmological constant is therefore $\Lambda=\frac{8\pi
G}{c^{4}}U=\frac{c^{3}}{\pi \hbar G}$, which can be rewritten as
$\Lambda_{\rm th }=\frac{1}{\pi}\frac{1}{r_{\rm Pl}^{2}}$, where
the Planck length is $r_{\rm Pl}=\sqrt{\frac{G\hbar}{c^{3}}}$.
However, it is also well known that the observed cosmological
constant can be expressed\footnote{This expression means that even
though $\Lambda$ is very small, it is comparable to the present
matter density.} as $\Lambda_{\rm ex}=\frac{3H^{2}}{c^{2}}\simeq
\frac{3}{r_{\rm Uni}^{2}}$, where $r_{\rm Uni}$ denotes the
cosmological radius (length scale of the universe). So, the ratio
of the theoretical cosmological constant, $\Lambda_{\rm th }$, to
the observed one, $\Lambda_{\rm ex}$, is
\begin{equation}
\frac{\Lambda_{\rm th }}{\Lambda_{\rm
ex}}=\frac{1}{3\pi}\left(\frac{r_{\rm Uni}}{r_{\rm
Pl}}\right)^{2}\simeq 10^{120},
\end{equation}
where the orders of magnitude of the Plank length and cosmological
radius ({\it i.e.}, $r_{\rm Pl}\simeq 10^{-35}$ m and $r_{\rm
Uni}\simeq 10^{26}$ m) have been inserted.

During the past 40 years, in an attempt to deal with the
cosmological constant problem, many theoretical works such as
adjustment mechanism, changing gravity, quantum cosmology and the
viewpoint of supersymmetry, supergravity and superstrings were
proposed\cite{Weinberg2}.  More recently, since some astrophysical
observations show that the large scale mean pressure of our
present universe is negative suggesting a positive cosmological
constant, and that the universe is therefore presently undergoing
an accelerating expansion\cite{Nature}, a large number of theories
and viewpoints (such as back reaction of cosmological
perturbations\cite{back}, QCD trace anomaly\cite{a}, contribution
of Kaluza-Klein modes to vacuum energy\cite{contribution},
five-dimensional unification of cosmological constant and photon
mass\cite{five}, nonlocal quantum gravity\cite{the}, quantum micro
structure of spacetime\cite{why}, relaxation of the cosmological
constant in a movable brane world\cite{relaxation}, the effect of
minimal length uncertainty relation (under the modified
commutation relation $[q, p]$) on the density of
states\cite{the2}, and so on) are put forward to resolve the
cosmological constant problem. In this paper, we will suggest a
so-called {\it comoving suppression mechanism} to resolve the
cosmological constant problem. Differing from the conventional
viewpoint that the observer is fixed at the comoving frame of
reference with the metric $g_{00}=1$, here we argue that the
observer may be located in a comoving coordinate system of the
metric $g_{00}=\frac{\lambda}{\Lambda}$ with $\lambda$ being a
certain parameter that will be discussed in the present paper. It
will be shown that in such a comoving frame of reference the
vacuum energy density can be greatly suppressed by the comoving
compression mechanism. In order to determine the comoving
parameter $\lambda$, we will make use of Mach's principle in the
discussion.

The paper discuss three topics: (i) the concept of $\lambda$- de
Sitter metric and the comoving parameter $\lambda$; (ii) the
calculation of vacuum energy density in the $\lambda$- de Sitter
universe; (iii) the varying observed speed of light and the
potential relation to the anomalous acceleration ($\sim
10^{-9}{\rm m/s}^{2}$) acquired by the Pioneer 10/11, Galileo and
Ulysses spacecrafts\cite{Anderson}.

\section{Comoving frame of reference and $\lambda$- de Sitter metric}
In this section, first we propose the concept of $\lambda$- de
Sitter world, the line element of which takes the form

\begin{equation}
 {\rm d}s^{2}=\frac{\lambda}{\Lambda}c^{2}{\rm
d}t^{2}-\exp \left(2\sqrt{\frac{\lambda}{3}}ct\right)[{\rm
d}r^{2}+r^{2}\left({\rm d}\theta^{2}+\sin^{2}\theta {\rm
d}\varphi^{2}\right)],           \label{element}
\end{equation}
which can be solved via Einstein field equation with the
cosmological constant term. In this comoving frame of reference
with $g_{00}=\frac{\lambda}{\Lambda}$ (where the coordinate time
is no longer the proper time), the cosmological expansion rate
equations are of the form
\begin{eqnarray}
& & \frac{8\pi G}{c^{2}}\frac{\lambda}{\Lambda}\rho+\lambda=\frac{\lambda}{\Lambda}\frac{3k}{R^{2}}+\frac{3\dot{R}^{2}}{c^{2}R^{2}},                  \nonumber \\
& & \frac{8\pi
G}{c^{4}}\frac{\lambda}{\Lambda}p=\lambda-\frac{\lambda}{\Lambda}\frac{k}{R^{2}}-\frac{\dot{R}^{2}}{c^{2}R^{2}}-\frac{2\ddot{R}}{c^{2}R},
\label{rate}
\end{eqnarray}
where $R$ denotes the cosmological scale factor. If we assume that
the Hubble parameter $H=\frac{\dot{R}}{R}$ is constant and we can
ignore the matter density and pressure\footnote{Recent observation
shows that the components of regular matter, dark matter and dark
energy in the universe is as follows: 4\% (regular matter), 23\%
(dark matter), 73\% (dark energy). This, therefore, means that the
component that is described by the cosmological constant term is
more than $\frac{2}{3}$ of the total matter in the whole
universe.} in Eq.(\ref{rate}), we may obtain
$\lambda=3\frac{H^{2}}{c^{2}}$. In this step, someone may think
that in Eq.(\ref{rate}) the comoving parameter $\lambda$ acts just
as the cosmological constant. This, however, may be not the true
case, the reason for which will be considered as follows: the
Hubble parameter $H$ was measured by the Doppler redshift
observation\footnote{The redshift is $z=\frac{\triangle
\lambda}{\lambda}$ with $\lambda$ being the wavelength of light.},
where the redshift
$z=\frac{\dot{R}}{R}\frac{L}{c\sqrt{\frac{\lambda}{\Lambda}}}$
with $L$ being the physical distance between the observed star and
the observer. Thus the observed Hubble parameter $H_{\rm
obs}=\frac{\dot{R}}{R}=zc\sqrt{\frac{\lambda}{\Lambda}}\frac{1}{L}$.
By using the above-obtained relation
$\lambda=3\frac{H^{2}}{c^{2}}$, one can arrive at
\begin{equation}
\Lambda=\frac{3z^{2}c^{2}}{L^{2}}.             \label{though}
\end{equation}
It should be noted that even though Eq.(\ref{though}) is a
familiar expression, it is of no help for overcoming the
cosmological constant problem, since here the speed of light $c$
is no longer the observed one. Instead, the observed speed of
light is $c'=c\sqrt{\frac{\lambda}{\Lambda}}$ (rather than $c$).
In the next section it will be verified that
$\sqrt{\frac{\lambda}{\Lambda}}=10^{-30}$. This, therefore,
implies that the fundamental constant $c$ is $10^{30}$ times more
than the observed speed of light ($c'=3.00\times 10^{8}$ m/s).
Thus it is believed that the above viewpoint cannot interpret the
cosmological constant problem.

The previous discussion shows that the {\it observed }
cosmological constant is not the physical meanings of $\lambda$,
namely, $\lambda$ cannot be considered a so-called {\it observed }
cosmological constant. The {\it observed } cosmological constant
is still $\Lambda$. In what follows we will consider the
compression mechanism of vacuum energy density in the $\lambda$-
de Sitter world ({\it i.e.}, the comoving frame of reference with
$g_{00}=\frac{\lambda}{\Lambda}$).

\section{Suppression of $\Lambda$ in $\lambda$- de Sitter world}
In this section, the vacuum energy density in the $\lambda$- de
Sitter world will be calculated and, by taking account of Mach's
principle, the reduction of $\Lambda$ by 120 orders of magnitude
will be demonstrated.

In the comoving coordinate system with
$g_{00}=\frac{\lambda}{\Lambda}$, the action of a particle is
$S=-mc\int {\rm d}s$ with
\begin{equation}
{\rm d}s=\sqrt{\frac{\lambda}{\Lambda}c^{2}{\rm d}t^{2}-\exp
\left(2\sqrt{\frac{\lambda}{3}}ct\right){\rm d}{\bf
r}^{2}}=\sqrt{\frac{\lambda}{\Lambda}c^{2}-v^{2}}{\rm d}t,
\end{equation}
where the test particle velocity squared is defined as $v^{2}=\exp
\left(2\sqrt{\frac{\lambda}{3}}ct\right)\left(\frac{{\rm d}{\bf
r}}{{\rm d}t}\right)^{2}$. Thus the Lagrangian of the test
particle under consideration is $
L=-mc\sqrt{\frac{\lambda}{\Lambda}c^{2}-v^{2}}$. It follows that
the canonical momentum and the canonical Hamiltonian read
\begin{equation}
 p=\frac{\partial L}{\partial
v}=\frac{mv}{\sqrt{\frac{\lambda}{\Lambda}-\frac{v^{2}}{c^{2}}}},
\quad
H=pv-L=\frac{m\frac{\lambda}{\Lambda}c^{2}}{\sqrt{\frac{\lambda}{\Lambda}-\frac{v^{2}}{c^{2}}}}.
\label{canonical}
\end{equation}
Let us define the observed mass and the observed speed of light as
follows: $m'=\frac{m}{\sqrt{\frac{\lambda}{\Lambda}}}$,
$c'=\sqrt{\frac{\lambda}{\Lambda}}c$. Clearly, the expressions
(\ref{canonical}) can be rewritten as
\begin{equation}
p=\frac{m'v}{\sqrt{1-\frac{v^{2}}{c'^{2}}}},   \qquad
H=\frac{m'c'^{2}}{\sqrt{1-\frac{v^{2}}{c'^{2}}}},
\end{equation}
which take the form familiar to us all. For the extremely
relativistic case $v\rightarrow c'$, we have $pc'\rightarrow H$. I
assume that the observers (say, we) are fixed actually in the
comoving frame of reference with $g_{00}=\frac{\lambda}{\Lambda}$,
and that the observed speed of light is $c'$ rather than $c$. The
calculated result for the vacuum energy density in the comoving
frame of reference with $g_{00}=\frac{\lambda}{\Lambda}$ is
\begin{eqnarray}
U&=&\frac{2}{(2\pi \hbar)^{3}}\int^{p_{\rm Pl}}4\pi
p^{2}\frac{1}{2}\hbar\omega_{p}{\rm d}p   \nonumber  \\
&\simeq &
\frac{2}{(2\pi \hbar)^{3}}\int^{p_{\rm Pl}}4\pi
p^{2}\frac{1}{2}(pc'){\rm d}p,
\end{eqnarray}
where the cutoff momentum is $p_{\rm Pl}=\sqrt{\frac{\hbar
c'^{3}}{G}}$. Note that here the fundamental constant $c$ has been
replaced with the observed speed of light $c'$ ($c'=3.00\times
10^{8}$ m/s). Further calculation yields $
U=\frac{c'^{7}}{8\pi^{2}\hbar G^{2}}$, and consequently the
calculated cosmological constant reads
\begin{equation}
\Lambda=\frac{8\pi G}{c^{4}}\frac{c'^{7}}{8\pi^{2}\hbar G^{2}}.
\label{cos}
\end{equation}
It should be noted that in Eq.(\ref{cos}) the coefficient
$\frac{8\pi G}{c^{4}}$ need not to be replaced with $\frac{8\pi
G}{c'^{4}}$, since here $c$ is a fundamental constant rather than
an observed speed of light. By using the relation ({\it i.e.},
$c=\sqrt{\frac{\Lambda}{\lambda}}c'$) between $c$ and $c'$,
Eq.(\ref{cos}) can be rewritten as $
\Lambda^{3}=\frac{{c'}^{3}}{\pi\hbar G}\lambda^{2}$. With the help
of $r_{\rm Pl}=\sqrt{\frac{G\hbar}{c'^{3}}}$, one can arrive at
\begin{equation}
\Lambda=\frac{\lambda^{\frac{2}{3}}}{\pi^{\frac{1}{3}}r_{\rm
Pl}^{\frac{2}{3}}},              \label{pl}
\end{equation}
which is expressed in terms of the Plank length and the comoving
parameter $\lambda$. Thus it follows from Eq.(\ref{pl}) that the
vacuum energy density (or cosmological constant $\Lambda$) is
determined only by the comoving parameter $\lambda$. If $\lambda$
is taken to be very small (or nearly vanishing), then the
corresponding $\Lambda$ is also very small. This is just the
comoving suppression mechanism for the cosmological constant. Even
though according to the conventional choice where it is assumed
that the observer is fixed in the comoving frame of reference with
$g_{00}=1$, the vacuum energy density is very large (even nearly
divergent), in the comoving frame with
$g_{00}=\frac{\lambda}{\Lambda}$, $\Lambda$ is greatly compressed
by many orders of magnitude.

Now the problem left to us is to determine the value of the
comoving parameter $\lambda$. Brief analysis indicates that it is
in connection with Mach's principle: specifically, substitution of
the observed cosmological constant $\Lambda\simeq \frac{3}{r_{\rm
Uni}^{2}}$ into Eq.(\ref{pl}) yields
\begin{equation}
\lambda=\frac{r_{\rm Pl}}{r_{\rm Uni}^{3}},     \qquad
\frac{\lambda}{\Lambda}\simeq \frac{r_{\rm Pl}}{r_{\rm Uni}}\sim
10^{-61}.                       \label{61}
\end{equation}
How can we understand the physical meanings of the obtained
$\lambda$ in Eq.(\ref{61})? The expression
$2\sqrt{\frac{\lambda}{3}}ct$ in the exponential factor of the
line element (\ref{element}) of the comoving frame with
$g_{00}=\frac{\lambda}{\Lambda}$ reads
\begin{equation}
2\sqrt{\frac{\lambda}{3}}cT=\frac{2}{\sqrt{3}}\frac{c'T}{r_{\rm
Uni}}={\mathcal O}(1),                  \label{o1}
\end{equation}
where the expression for $c$, {\it i.e.},
$c=\frac{c'}{\sqrt{\frac{\lambda}{\Lambda}}}\sim 10^{38}{\rm m/s}$
has been inserted. In Eq.(\ref{o1}) $T$ denotes the cosmological
age of the present universe. In accordance with the result in
Eq.(\ref{o1}), one can conclude without fear that the comoving
parameter $\lambda$ taking the form (\ref{61}) just agrees with
Mach's principle.

Additionally, it should be pointed out that the observed (and
measured) mass $m'$ of a particle is not the real mass $m$, the
relation between which is given
$m=\sqrt{\frac{\lambda}{\Lambda}}m'\simeq 10^{-30}m'$. It may be
believed that this relation also has close relation to Mach's
principle which holds that the inertial properties of a body are
determined by the mass distribution in the universe, and the
inertial force acting upon a body arises from an interaction
between it and the matter-energy content of the whole universe. It
is of physical interest that the real mass $m$ is actually merely
one part of $10^{30}$ of the observed one, which, therefore, means
that the real total cosmological mass is only about $10^{23}$ Kg
that is surprisingly less than that of the Earth (the observed
mass of the Earth is $6.0\times 10^{24}$ Kg)!

In order to close this section, we briefly conclude the comoving
suppression mechanism for the cosmological constant with some
remarks. In view of the above discussion, the comoving suppression
mechanism is established based on the following four points: (i)
the assumption that the observer is fixed in a comoving frame of
reference with $g_{00}=\frac{\lambda}{\Lambda}$; (ii) Mach's
principle that can determine the comoving parameter $\lambda$;
(iii) the observed speed of light
$c'=c\sqrt{\frac{\lambda}{\Lambda}}$; (iv) the vacuum energy
density $U=\frac{c'^{7}}{8\pi^{2}\hbar G^{2}}$ in the comoving
frame of reference with $g_{00}=\frac{\lambda}{\Lambda}$. Thus the
comoving suppression mechanism will inevitably lead to the
relaxation of the cosmological constant by 120 orders of
magnitude.

\section{The varying observed speed of light}
According to the above comoving suppression mechanism, it is
readily verified that the observed speed of light $c'$ is not
constant. In fact, its variation (rate of change) at present is
\begin{equation}
\frac{{\rm d}c}{{\rm d}t}={\mathcal O}\left(10^{-9}{\rm
m/s}^{2}\right).
\end{equation}

In 1998, Anderson {\it et al.} reported that, by ruling out a
number of potential causes, radio metric data from the Pioneer
10/11, Galileo and Ulysses spacecraft indicate an apparent
anomalous, constant, acceleration
acting on the spacecraft with a magnitude $\sim 8.5\times $ $10^{-8}$cm/s$%
^{2}$ directed towards the Sun\cite{Anderson}. Is it the effects
of dark matter or a modification of gravity? Unfortunately,
neither easily works. It is interesting that by taking the cosmic
mass, $M=10^{53}$ kg, and cosmic scale, $R=10^{26}$ m, our
calculation shows that {\it this acceleration is just equal to the
value of field strength on the cosmic boundary due to the total
cosmic mass}. This fact leads us to consider a theoretical
mechanism to interpret this anomalous phenomenon. The author
favors that the gravitational Meissner effect may serve as a
possible interpretation. Here we give a rough analysis, which
contains only the most important features rather than the precise
details of this theoretical explanation. Parallel to London$^{,}$s
electrodynamics of superconductivity, it shows that gravitational
field may give rise to an {\it effective rest mass
}$m_{g}=\frac{\hbar }{c^{2}}\sqrt{8\pi G\rho _{m}}$ due to the
self-induced charge current \cite{Hou}, where $\rho _{m}$ is the
mass
density of the universe. Then one can obtain that $\frac{\hbar }{m_{g}c}%
\simeq 10^{26}$m that approximately equals $R$, where the mass
density of the universe is taken to be $\rho _{m}=0.3\rho
_{c}$\cite{Pea} with $\rho _{c}\simeq 2\times 10^{-26}$kg/m$^{3}$
being the critical mass density. An added constant acceleration,
$a$, may result from the Yukawa potential and can be written
as\footnote{It may also be calculated as follows: $a =\frac{GM}{2}\left( \frac{m_{g}c}{\hbar }\right) ^{2}=\frac{GM}{c^{2}}%
(4\pi \rho _{m}G)=\frac{GM}{c^{2}R}\frac{G(4\pi R^{3}\rho _{m})}{R^{2}}\simeq \frac{GM}{%
R^{2}}$, where use is made of $\frac{GM}{c^{2}R}\simeq {\mathcal
O}(1),4\pi R^{3}\rho _{m}\simeq M$, which holds when the
approximate evaluation is performed.}

\begin{equation}
a=\frac{GM}{2}\left( \frac{m_{g}c}{\hbar }\right) ^{2}\simeq
\frac{GM}{R^{2}}\simeq  10^{-9}{\rm m/s}^{2}.
 \label{eq31}
\end{equation}

Hence we demonstrated that the value of $10^{-9}{\rm m/s}^{2}$ is
a very typical and interesting ``acceleration'', which arises in
(i) the anomalous acceleration acting on the Pioneer 10/11,
Galileo and Ulysses spacecrafts\cite{Anderson}, (ii) gravitational
Meissner effect\cite{Meissner}, (iii) the gravitational field
strength at the cosmic boundary, and (iv) even the varying
observed speed of light. It is reasonably believed that these
facts may clue physicists on the mystery of anomalous acceleration
acquired by the Pioneer 10/11, Galileo and Ulysses
spacecrafts\cite{Anderson}. Since this subject is beyond the scope
of the present paper, we will not consider it further.

\section{Concluding remarks}
The comoving suppression mechanism for the cosmological constant
problem assumes that we (observers) are fixed in a comoving
coordinate system with $g_{00}=\frac{\lambda}{\Lambda}$ rather
than with $g_{00}=1$, and that the comoving parameter $\lambda$
can be determined by Mach's principle. In such a theoretical
framework, the calculation of the vacuum energy density in this
comoving frame of reference indicates that the cosmological
constant is reduced by 120 orders of magnitude.

In the conventional comoving frame of reference, the metric of
de-Sitter world is

\begin{equation}
 {\rm d}s^{2}=c^{2}{\rm
d}t^{2}-\exp \left(2\sqrt{\frac{\Lambda}{3}}ct\right){\rm d}{\bf
r}^{2}
\end{equation}
rather than Eq.(\ref{element}), {\it i.e.},
\begin{equation}
 {\rm d}s^{2}=\frac{\lambda}{\Lambda}c^{2}{\rm
d}t^{2}-\exp \left(2\sqrt{\frac{\lambda}{3}}ct\right){\rm d}{\bf
r}^{2}.           \label{conclud}
\end{equation}
It is believed that the comoving parameter $\lambda$ in
(\ref{conclud}) possesses physical meanings and therefore deserves
essential consideration. Note that since in Eq.(\ref{conclud}),
$\frac{\lambda}{\Lambda}c^{2}=c'^{2}$ with $c'=3.00\times 10^{8}$
m/s, our experimental measurements (astrophysical observations)
cannot demonstrate whether we are just in a comoving frame of
reference with $g_{00}=\frac{\lambda}{\Lambda}$ rather than with
$g_{00}=1$, which, therefore, means that the comoving parameter
$\lambda$ should be determined on by Mach's principle. But, what
is the physical origin of Mach's principle? This problem seems to
be of no satisfactory solutions up to now. So, the key problem of
the comoving suppression mechanism for the cosmological constant
problem in the present paper may be the resolution of the physical
meanings and origin of Mach's principle, which is now under
consideration and will be published elsewhere.

$^\ast${\bf Shen's electronic address}: jqshen@coer.zju.edu.cn


\begin{references}
\bibitem{history} on the history of the cosmological constant
problem, see, for example, Straumann, N., (2002). gr-qc/0208027.

\bibitem{Weinberg} Weinberg, S., (2000). astro-ph/0005265.

\bibitem{Weinberg2}   Weinberg, S., (1988). {\it Rev. Mod. Phys.}
{\bf 61}, 1 and references therein.

\bibitem{Nature} Perlmutter, S. {\it et al.}, (1998).
{\sl Nature} {\bf 391,} 51; Perlmutter, S. {\it et al.}, (1999).
{\sl Astrophys. J.} {\bf 517,} 565; Riess, A.G. {\it et al.},
(1998). {\sl Astrophys. J.} {\bf 116,} 1009.

\bibitem{back} Brandenberger, R.H., (2002). hep-th/0210165.

\bibitem{a}  Sch\"{u}tzhold, R., (2002). gr-qc/0204018.

\bibitem{contribution} Gupta, A., (2002). hep-th/0210069.

\bibitem{five} Kohler, C., (2002). gr-qc/0202076.

\bibitem{the} Moffat, J.W., (2002). hep-th/0207198;
hep-th/0210140.

\bibitem{why}   Padmanabhan, T., (2002). gr-qc/0204020.

\bibitem{relaxation} Khlebnikov, S., (2002). hep-th/0207258.

\bibitem{the2} Chang, L.N., Minic, D., Okamura, N., and Takeuchi,
T., (2002). hep-th/0201017.

\bibitem{Anderson}  Anderson, J.D., Laing, P.A., Lau, E.L., {\it et al.}, (1998).
{\sl Phys. Rev. Lett.} {\bf 81,} 2858.

\bibitem{Hou}  Hou, B.Y. and Hou, B.Y., (1979). {\sl Phys. Ener. Fort.
Phys. Nucl.} {\bf 3,} 255.

\bibitem{Pea}  Peacock, J.A., Cole, S., Norberg, P., {\it et al.}, (2001). {\sl %
Nature} {\bf 410,} 169.

\bibitem{Meissner} Shen, J.Q., (2003). gr-qc/0305094.
\end{references}
\end{document}